# Silicon Waveguides and Ring Resonators at 5.5 μm

Alexander Spott[1], Yang Liu[1], Tom Baehr-Jones[1], Rob Ilic[2], and Michael Hochberg[1]

[1]*Department of Electrical Engineering, University of Washington, Campus Box 352500, Seattle WA 98195-2500 USA*

[2]*Cornell NanoScale Science and Technology Facility, Cornell University, Ithaca NY 14853 USA*

We demonstrate low loss ridge waveguides and the first ring resonators for the mid-infrared, for wavelengths ranging from 5.4 to 5.6 μm. Structures were fabricated using electron-beam lithography on the silicon-on-sapphire material system. Waveguide losses of 4.0±0.7 dB/cm are achieved, as well as Q-values of 3.0 k.

Silicon waveguides have been primarily operated at wavelengths in the near-infrared (NIR), typically around 1.4-1.6 μm. This has been convenient due to the large number of commercial optical components available in this regime. The mid-infrared (MIR) wavelengths, typically defined to range from 2-20 μm[1], have proven to be useful for a number of applications. Many astronomy experiments depend upon the detection of MIR wavelengths[2]. Chemical bond spectroscopy benefits from a large range of wavelengths from visible to past 20 μm[3]. Thermal imaging (such as night vision) depends upon mid-infrared wavelengths as a source of blackbody radiation[4].

Silicon waveguides for MIR wavelengths have previously been theorized and fabricated[5]; in fact, an optical parametric amplifier at 2.2 μm[6] was recently shown in silicon waveguides. But there had not been results at longer wavelengths until our recent demonstration of waveguides on a silicon-on-sapphire (SOS) substrate for wavelengths near 4.5 μm[7]. With low-loss waveguides, it is possible to construct high-Q ring resonators. Ring resonators in the near-infrared have had a number of applications, including biosensing[8], modulation[9] and wavelength conversion[10]. Here we show SOS based waveguides and the first ring resonators at wavelengths between 5.4-5.6 μm. We demonstrate a waveguide loss of 4.0±0.7 dB/cm and a ring resonator Q-value of 3.0 k.

We fabricated ridge waveguides with dimensions 1.8 x 0.6 μm on an SOS substrate. The mode structure and method for modesolving is described in our previous work. Near 5.5 um wavelengths, only the TE0 mode should guide[7]. The ridge waveguides were fabricated on a 100 mm diameter epitaxial SOS wafer fragment with an electron-beam lithography system using standard maskless lithography techniques[11].

Our chip contained a variety of both simple ridge waveguides and ring resonators. The simple waveguides were designed to have a number of different lengths in order to characterize waveguide loss. Two of these guides were designed with dimensions of 1.2 x 0.6 μm as control structures, and failed to achieve guiding as predicted. Ring resonators were fabricated with a variety of different radii and edge-to-edge spacing (coupling spacing from waveguide to ring). The primary Q-value result of 3.0 k was obtained from a ring with a 40 μm radius and 0.25 μm edge-to-edge spacing. A micrograph of this ring is shown in figure 1.

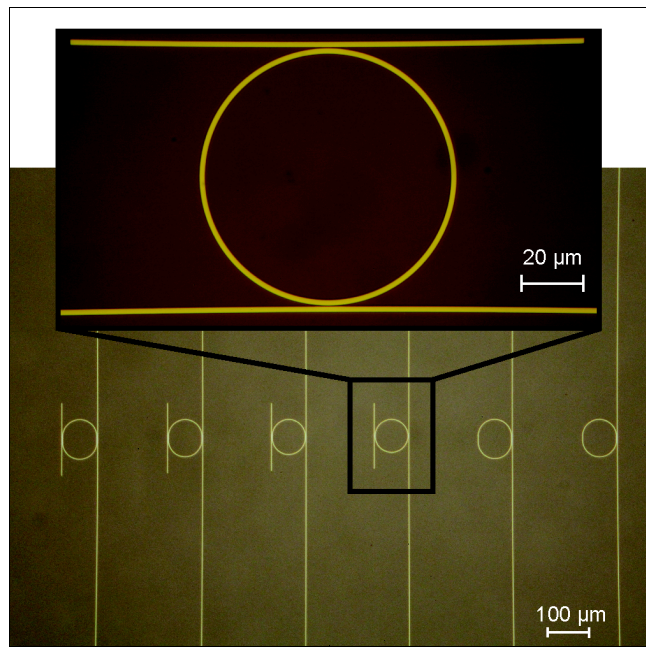

Figure 1: Optical micrographs of the primary ring resonator device found to have a Q-factor of 3.0 k (top) and a group of ring resonators of various dimensions (bottom).

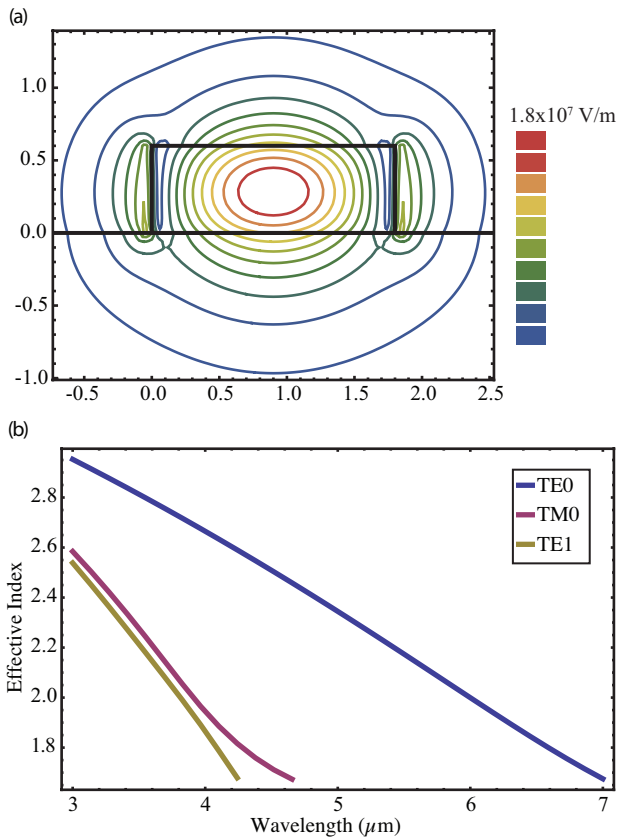

Figure 2: (a) A contour plot of the optical mode pattern of the waveguide. Dimensions are in μm. (b) A dispersion diagram for SOS waveguides.

All waveguides tested terminated on both ends with a wider 8.0 x 0.6 μm waveguide in order to improve edge coupling efficiency. These wider sections were roughly 5 mm long (before cleaving) and were connected to the main guides with a simple taper. The chip was cleaved manually through these runouts. Approximately half of the 5 mm long section remained on each side of the devices after cleaving. Since the length of this portion of the device was identical for all devices tested, it did not affect the waveguide loss measurement, as is shown later.

Figure 3 shows a diagram of our test setup. We achieved a total dynamic range near 85 dB and attained an insertion loss of around 25 dB coupling into the waveguides[12].

The possibility of TM guiding was also investigated by rotating the input polarization 90 degrees. We found a signal at least 15 dB lower than for the TE input, which was close to the noise floor. We suspect the source of the small signal seen is simply scattering off the waveguide endfacets, or possibly due to a very small amount of TE light remaining in the system.

Measured waveguide loss was 4.0±0.7 dB/cm. Waveguide loss was determined by a least-squares linear regression of transmission power from 8 separate guides of varying lengths between 4.1 and 16.4 mm. Primary loss measurements were taken with the laser operating around 100 mW and 5.5 μm. Similar losses were seen with the laser operating near 6 mW, suggesting that there is minimal nonlinear loss.

Wavelength transmission spectra were taken on multiple ring resonators. We focused measurements on a ring with a radius of 40 μm, edge-to-edge spacing of 0.25 μm, and a drop port. Adding a drop port will lower the Q value, but make it easier to obtain a large extinction when on resonance. This ring yielded a Q-value of 3.0 k, a free spectral range (FSR) of roughly 29.7 nm, and an associated group index of 3.99. Without prior knowledge of the waveguide loss, it was difficult to optimally design the ring resonators. At a waveguide loss of 4.0 dB/cm, it would theoretically be possible to achieve a critically coupled Q-value as high as 25 k. This discrepancy is almost certainly due to the presence of the drop port, and we expect to greatly improve these numbers in the future. The theoretical FSR of a ring of this size is 29.2 nm (with a theoretical group index of 4.05), which is in close agreement with our typical measured FSR.

Atmospheric absorption becomes an increasing problem at longer wavelengths in the mid-infrared. For most of our measurements, we simply performed testing at a wavelength where there was no observable absorption from the atmosphere. We also conducted further wavelength sweeps of the primary ring result in a nitrogen-purged environment, in order to obtain a cleaner transmission spectrum. We found that all absorption peaks were removed, giving a much clearer indication of resonator peaks.

In conclusion, we have shown that it is possible to fabricate low loss ring resonator devices with Q-values as high as 3.0 k on a SOS substrate for wavelengths from 5.4 – 5.6 μm. We expect it will be possible to improve the Q-values in the near future, due to the low waveguide loss measured.

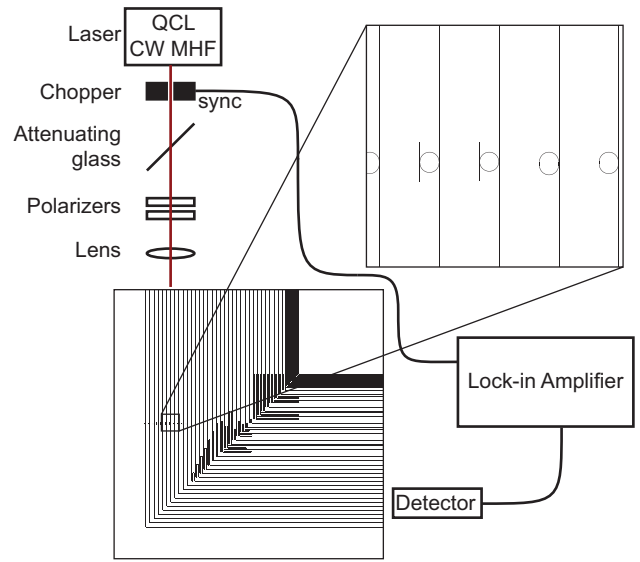

Figure 3: A diagram of the test setup including an image of the chip tested.

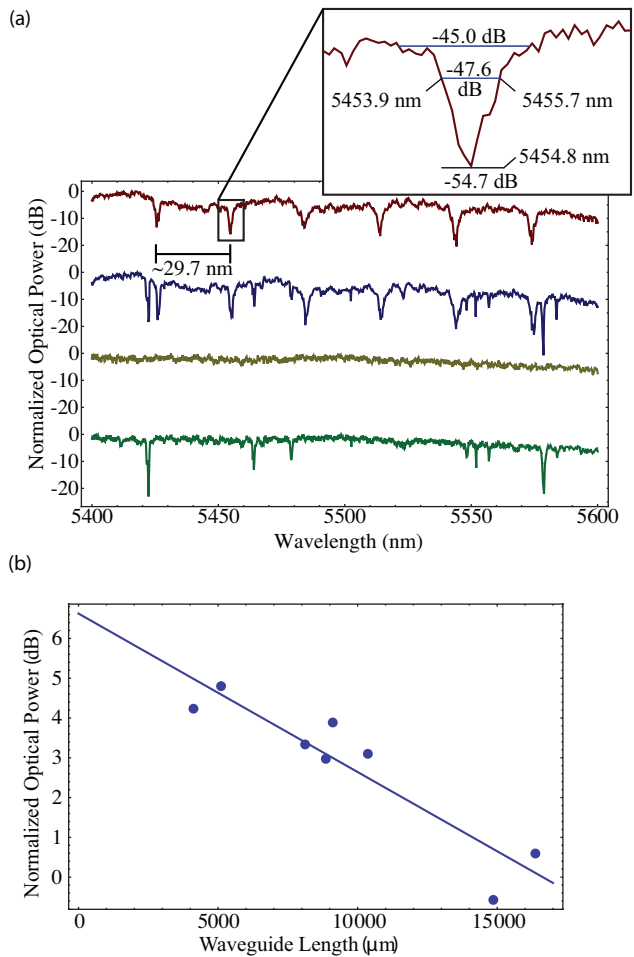

Figure 4: (a) Transmission spectra for the primary ring reported: tested in a nitrogen-purged environment (top) and tested under normal atmospheric conditions (second). Transmission spectra for a regular waveguide: tested under a nitrogen-purged environment (third) and tested under normal atmospheric conditions (bottom). A Q-factor of 3.0 k and an FSR of 29.7 can be seen. (b) A plot of transmitted optical power for various devices of a variety of waveguide lengths. A least-squares linear approximation is shown, demonstrating waveguide losses of 4.0±0.7 dB/cm.

The authors would like to thank Gernot Pomrenke, of the Air Force Office of Scientific Research, for his support through AFOSR YIP and PECASE awards, and would like to acknowledge support from the NSF STC MDITR Center, the Washington Research Foundation, the Mary Gates Endowment, and the Washington NASA Space Grant Consortium.

# Supplemental Materials

## Fabrication details

Hydrogen Silsesquioxane (HSQ) resist was first spun and baked. In order to improve the selectivity between Si and resist, the HSQ was hardened with $O_2$ plasma by a Branson Barrel Asher. The resist was then exposed with a base dose of 800 $\mu C/cm^2$ at a beam current of 5 nA, as well as 64 doses of proximity correction. Etching was achieved with PT770. Finally, remaining HSQ needed to be stripped with an HF wash.

## Test setup details

Waveguides were tested with a Daylight Solutions 21052-MHF-024-D0017 tunable, mode-hop free, quantum cascade laser, able to emit TE polarized mid-infrared wavelengths between 5.18 and 5.65 µm. The light was first attenuated by 10 dB through a 0.15 mm thin glass microscope slide angled near 45 degrees to the laser in order to minimize the influence of back-reflection on the laser cavity. The beam was then coupled through two $BaF_2$ wire grid polarizers in order to ensure that the TE polarization was kept constant, and allow us to produce TM polarized light. The beam was roughly 1.5 mm wide and 1 mm tall and a ZnSe lens focused it down into a spot size near 60 µm wide in order to facilitate edge coupling with the waveguides. The signal-to-noise ratio was dramatically improved by chopping the beam at ~6 kHz, then locking into that frequency with a lock-in amplifier. With a detector responsivity of roughly 1.2 A/W we achieved a total dynamic range near 85 dB.

A 50 µm aperture was used in front of the lens to locate the position of the focal point, making it possible to identify which guide was being coupled to. Once guiding was achieved we removed the aperture in order to improve the signal.